\documentclass[prc,aps,a4paper,groupedaddress,superscriptaddress,nofootinbib,showpacs
,preprintnumbers,onecolumn,12pt]{revtex4-1}
\def\e{{\rm e}}

\usepackage[english]{babel}
\usepackage[utf8]{inputenc}
\usepackage{amssymb,ulem}
\usepackage{amsmath}
\usepackage{indentfirst}
\usepackage{graphicx}
\usepackage{subfigure}
\usepackage{multirow}
\usepackage{color}

\newcommand{\mathsym}[1]{{}}

\def\e{{\mathrm{e}}}

\def\betab{b}

\def\d{{\rm d}}
\def\e{{\rm e}}

\def\bfr{{\bf r}}

\def\bfalpha{\boldsymbol{\alpha}}
\def\bfsigma{\boldsymbol{\sigma}}
\def\la{\lambda}

\def\aa{a}
\def\bb{b}

\begin{document}
\title{Neutron stars with Bogoliubov quark-meson coupling model \footnote{This paper is dedicated to the memory of our friend and colleague Steven Moskowsky, to which he gave a crucial contribution.}}
\author{Aziz Rabhi} 
\email{rabhi@uc.pt}
\affiliation{University of Carthage, Avenue de la R\'epublique BP 77 -1054 Amilcar, Tunisia,}
\affiliation{CFisUC, Department of Physics, University of Coimbra, 3004-516 Coimbra, Portugal.}

\author{Constan\c ca Provid\^encia}
\email{cp@uc.pt}
\affiliation{CFisUC, Department of Physics, University of Coimbra, 3004-516 Coimbra, Portugal.}

\author{Steven A. Moszkowski}
\email{deceased}
\affiliation{UCLA, Los Angeles, CA 90095, USA}

\author{Jo\~ao da Provid\^encia}
\email{providencia@uc.pt}
\affiliation{CFisUC, Department of Physics, University of Coimbra, 3004-516 Coimbra, Portugal.}

\author{Henrik Bohr}
\email{hbohr@dtu.dk}
\affiliation{Department of Physics, B.307, Danish Technical University, DK-2800 Lyngby, Denmark}
\date{\today}
\begin{abstract}
A quark-meson coupling model based on the quark model proposed by Bogoliubov
for the description of the quark dynamics is developed and applied to
the description of neutron stars. Starting from a SU(3) symmetry
approach, it is shown that this symmetry has to be broken in order to
satisfy  the constraints set by the hypernuclei and by neutron stars.  The model is able to
describe observations such as two solar mass stars or the radius of
canonical neutron stars within the uncertainties presently accepted.
If the optical potentials for $\Lambda$ and $\Xi$ hyperons in
symmetric nuclear matter at saturation  obtained from laboratory measurements of
hypernuclei properties are  imposed the model predicts no strangeness inside
neutron stars.
\end{abstract}
\pacs{26.60.-c, 26.60.+c, 21.65.+f, 97.60.Jd}
\maketitle

\section{Introdution}
The study of nuclear matter properties has received, in the recent few decades, much attention.
Such investigations are particularly important in connection with nuclear-astrophysics.
The recent  detection of the gravitational waves  GW170817 and 
  the follow-up of  the electromagnetic counterpart from a
neutron star merger \cite{abbott17,abbott18,multim17}, together with the simulataneous measurement of
the radius and mass of the pulsar PSR  J0030-0451 by NICER \cite{nicer1,nicer2}
are very important observations to constrain the equation of  state
(EoS) of  dense matter.  Besides, the two solar mass pulsars PSR
J1614-2230 \cite{demorest,j1614a}, PSR J0348+0432
\cite{Antoniadis2013} and MSP J0740+6620 \cite{cromartie2020}  are also setting
important constraints on the nuclear matter EoS at high
densities. In particular, these large masses put some difficulties on
the possible existence of non-nucleonic degrees of freedom, such as hyperons
or quark matter, in the inner core of the NS. In \cite{demorest}, it was even suggested that  PSR
J1614-2230 would rule out the appearance of these degrees of freedom inside pulsars. Since then many works
have shown that the present  existing  constraints on the high density
EoS are suffciently weak to still allow for the onset of hyperons, quarks or
other non-nucleonic degrees of freedom inside two solar mass neutron
stars \cite{bednarek2012,chatterjee1,chatterjee2,colucci2013,providencia2013,colucci2014,fortin2016}.

In relativistic mean field (RMF) models \cite{walecka,boguta,muller,typel},
the nucleon-nucleon interaction is described in terms of the coupling of nucleons,
assumed to be point particles, with isoscalar
scalar mesons, isoscalar vector mesons, and  isovector vector mesons. In
order to describe adequately nuclear matter properties RMF models
include either  self- and cross-interactions  \cite{boguta,muller}
among these mesons or 
density dependent couplings \cite{typel}.
There have been attempts, based on the MIT bag model  \cite{guichon}, and on the Nambu-Jona-Lasinio
(NJL) model \cite{bentz}, to take into account the quark structure of the nucleon, in order to
incorporate  the 
meson couplings at a more basic level.
Along these lines, the nuclear equation of state (EoS) has been obtained
and the properties of nuclear matter have been determined
by Guichon, Saito and Thomas \cite{guichon, guichon1,tsushima98,
guichon2}, and by others \cite{qmcPanda, qmcPanda1,panda12},
in the framework of quark-meson coupling (QMC) models. 
Still within the same
model, in \cite{shen1,shen2} the authors have studied the effect of strong magnetic fields on kaon condensation and
hyperonic matter, respectively.
  An improved version of the QMC model based on the MIT bag model, which includes the
  polarizability of the nucleons and uses Hartree-Fock, has been
  proposed in \cite{guichon2,stone07,stone18}.
More recently,
nuclear matter
has also been investigated
in the context of a modified QMC model based on the
replacement of the nucleon bag by an independent quark potential
\cite{batista,barik1,barik2,barik3}. In this model confinement is described by  including
 an  equally mixed scalar-vector harmonic  potential and extra
  corrections that take into account chiral symmetry restoration and
  gluon exchange.

In QMC models the hyperon parameters are fixed to the hyperon masses
  \cite{tsushima98,qmcPanda1,panda12,stone07,stone18}. The coupling of
  the $\sigma$-meson to hyperons is obtained selfconsistently, while
  the $\omega$-hyperon coupling is fixed by symmetry arguments or by
  imposing that  accepted values of the hyperon
  potential in symmetric nuclear matter at saturation are
  reproduced. This last procedure is the one that has also been
  adopted in \cite{barik3} for the MQMC model. In this model the
  incompressibility depends on the quark mass and only taking a quark
  mass above 200 MeV, is it possible to describe neutron
  stars containing hyperons in their core with masses above 1.9
  $M_\odot$. In \cite{stone07,stone18} the use of an
  Hartree-Fock approach and the explicit introduction of a
  polarizability term in the hyperon description pushes the onset of
  hyperons to densities close to 4 $\rho_0$.  Both  in the above QMC
  and MQMC models the $\sigma,\, \omega, \, \rho$-mesons  couple
  exclusively to the  $u$ and $d$ quarks, corresponding to a perfect $\omega-\phi$
  mixing.

Motivated by the idea of the string tension,
Bogoliubov proposed an independent quark model
for the description of the quark dynamics
\cite{bogolubov}.
The phenomenological description of
hadronic matter  in the spirit of the QMC approach, combined with
Bogoliubov's interesting quark model, has been considered in \cite{bohr}, for non-strange matter, and in \cite{panda},
for strange matter.
We will refer to the model considered in \cite{bohr,panda} as the
Bogoliubov-QMC model.
This model assumes that baryons are composed of quarks bound
by a linearly raising potential, as suggested by gauge theories.  The constituent mass of the $u,\, d$ quarks is generated
dynamically. 
In the present study,  we consider  a generalization  of  the model
proposed in \cite{panda}, where the couplings of the quarks $s$ to
vector bosons have not been explicitly considered. Instead, in
\cite{panda} it is postulated that the couplings of hyperons to the
vector mesons 
are well constrained by the phenomenological hyperon potentials in
nuclear matter, a procedure similar to the one undertaken in \cite{panda12,barik3}.
In the following, the consequences of considering the coupling of the quarks
$u,\,d,\,s$ to appropriate vector bosons starting from a SU(3)
symmetry approach  are explicitly investigated.
We discuss under  which conditions it is possible to describe two solar
mass  stars with a non zero strangeness content and determine their
chemical content. In particular, we conclude that this symmetry has to be broken in order to satisfy the constraints set by the hypernuclei and by neutron stars.

In section II we briefly present the model, in section III the
description of  hadronic matter with  strangeness is introduced  and
the 
$\beta$-equilirium equation of state is built. 
In the section IV, we  obtain  the structure  and properties of
neutron stars described by the present models and  discuss the
results. Finally some concluding remarks  are drawn in the last  section.

\color{black}
\section{The model}
We consider the Hamiltonian
\begin{equation}
\label{h_D}
h_D=-i\bfalpha\cdot\nabla+\beta\left(\kappa|\bfr|+m-g^q_\sigma
  \sigma\right).
\end{equation}
Here, $m$ is the current quark mass, $\beta$ and the components $\alpha_x,\alpha_y,\alpha_z$ of
$\bfalpha$ are Dirac matrices, $\sigma$ denotes the external scalar
field, $g^q_\sigma$ denotes the coupling of the quark to the $\sigma$ field and $\kappa$
denotes the string tension,
$$
\beta=\left[\begin{matrix}I&0\\0&-I\end{matrix}\right],~~
\alpha_x=\left[\begin{matrix}0&\sigma_x\\\sigma_x&0\end{matrix}\right],~~
\alpha_y=\left[\begin{matrix}0&\sigma_y\\\sigma_y&0\end{matrix}\right],~~
\alpha_z=\left[\begin{matrix}0&\sigma_z\\\sigma_z&0\end{matrix}\right],~~
$$
where $\sigma_x,~\sigma_y,~\sigma_z$ are the Pauli matrices.
The current quark mass  $m$ is taken to be
$m=0$ for $u,d$ quarks because their constituent mass
is assumed to be determined exclusively by the value of $\kappa$. 
The constituent mass of the $u,\, d$ quarks is generated dymamically,
while the constituent mass of the $s$ quark arises both dynamically and
from its ``current'' mass. Considering  a SU(3) symmetry,
the coupling $g^q_{\sigma}$ is assumed the same for quarks $u,\,  d,\,
s$.
The eigenvalues of $h_D$ are obtained by a scale transformation from the eigenvalues of
\begin{equation*}h_{D_0}=-i\bfalpha\cdot\nabla+\beta\left(|\bfr|-\aa\right).
\end{equation*}
The parameter $a$ is related to the nucleon radius. We need the lowest positive eigenvalue of $h_{D_0}$.
We cannot apply the variational principle to $h_{D_0}$, because its eigenvalues are not bounded from below, but we can apply
the variational principle to the square of the Hamiltonian,
\begin{equation}\label{2D0} h_{D_0}^2=-\nabla^2+(|\bfr|-
a)^2+i\beta\bfalpha\cdot{\bfr\over|\bfr|}.\end{equation}
We wish to determine variationally the lowest positive eigenvalue of $h_{D_0}$ versus $a$.
The variational ansatz should take into account the Dirac
structure of the quark wave-function,
so that we consider the following ansatz,
\begin{equation}\label{newansatz}\Psi_{\betab,\lambda}=\left[\begin{matrix}\chi
\\i\lambda(\bfsigma\cdot\bfr)\chi\end{matrix}\right]\e^{-(|r|-\aa-\bb)^2/2},\end{equation}
where ${\betab,\lambda}$ are variational parameters, and $\chi$ is a
2-spinor. From the wave function in Eq.~(\ref{newansatz}), it is clear
that $a + b$ is a measure of the nucleon radius. Minimizing the expectation value of $h_{D0}^2$ for
$\Psi_{\betab,\lambda},$ the following expression for the quark mass
is found,
\begin{equation}{m^2(\kappa,\aa)\over\kappa}=\min_{\lambda,\bb}{\langle \psi_{\betab,\lambda}|h_{D0}^2|\psi_{\betab,\lambda}\rangle
\over\kappa\langle
\psi_{\betab,\lambda}|\psi_{\betab,\lambda}\rangle}=
\min_{\lambda,\bb}{{\cal K}_0+{\cal V}_0+{\cal V}_{01}\la+({\cal
K}_1+{\cal V}_1)\la^2\over {\cal N}_0+{\cal
N}_1\la^2},\label{m^2/K}\end{equation} 
where ${\cal N}_0$, ${\cal V}_0$, ${\cal K}_0$, ${\cal N}_1$, ${\cal V}_1$, and ${\cal K}_1 $ are all given in~\cite{bohr}.

Minimization of Eq. (\ref{m^2/K}) with respect to $\lambda$ is readily
performed, so that
\begin{equation}{m^2(\kappa,\aa)\over\kappa}={1\over2}\min_\betab\left({{\cal K}_0+{\cal V}_0\over{\cal N}_0}+{{\cal K}_1+{\cal V}_1\over{\cal N}_1}
-\sqrt{\left({{\cal K}_0+{\cal V}_0\over{\cal N}_0}-{{\cal
K}_1+{\cal V}_1\over{\cal N}_1}\right)^2+\left({{\cal
V}_{01}\over\sqrt{{\cal N}_0{\cal
N}_1}}\right)^2}~\right)\label{m^2-alpha}.\end{equation}

Minimization of the r.h.s. of Eq. (\ref{m^2-alpha}) with respect to
$\bb$ may be easily implemented.
We have found that in the interval $-1.25<a<2.4$, that covers the
range of densities we will consider, we may express the groundstate energy, $m(\kappa,\aa),$  of $h_{D0}$,
with sufficient accuracy, as
\begin{eqnarray}\label{quarkmass}
\frac{m(\kappa, a)^2}{\kappa}&=&
2.64123
-2.35426 a
+0.825225 a^2
-0.072244 a^3 \cr
&&-0.0314736 a^4
+0.00155171 a^5
   +0.00257144 a^6.
\end{eqnarray}
Taking $a=g^q_\sigma\sigma/\sqrt{\kappa}$ for quarks $u,\,d$, we get, in the vacuum, the constituent mass of these quarks  equal to 313 MeV, with $a=0$ and $\kappa=37106.931784$ MeV$^2$. For the quark $s$, $a=a_s=-1.2455+g^q_\sigma\sigma/\sqrt{\kappa}$ reproduces the  vacuum  constituent mass 504 MeV of this quark. Consequently, the mass $M_{B}^*$ of the baryon $B$ is given as follows
\begin{equation}
M^*_N=M^*_P=3 m(\kappa, a), M^*_\Lambda= 2 m(\kappa, a)+m(\kappa,
a_s), M^*_\Xi= m(\kappa, a)+2 m(\kappa, a_s).
\label{MB}
\end{equation}
As we will discuss in the following, the $\Sigma$-hyperons will not be considered, because experimental data seem to indicate that the potential of  the  $\Sigma$-hyperon in nuclear matter is quite repulsive \cite{gal}, so that their appearance is disfavored.

\section{Hadronic matter}

In order to describe hadronic matter, we introduce the vector-isoscalar $\omega$ meson, the vector-isovector $b_3$ meson and use nuclear matter properties to fix the couplings of these mesons to nucleons.

In the present model, the field $\omega$  is replaced by a vector field of the $\eta$ type, in the spirit of the reference \cite{glashow}, 
with structure $(\bar u u+\bar d d+(1+\delta)~\bar s s )/\sqrt{2+(1+\delta)^2}$, where $1+\delta>0,$
so that the coupling of the $\omega$-meson to the quark $s$ is equal to the coupling to the
quarks $u,~d$ multiplied by $1+\delta$. This {\it ansatz} breaks the SU(3) symmetry and, as will be shown later, stiffens the EoS.
The parameter $\delta$  will be fixed by the
potencial $U_\Lambda$ of the $\Lambda$-hyperon in symmetric nuclear
matter at saturation, and  by imposing the existence of two solar mass stars.

In this framework, the energy density
is given by
\begin{eqnarray}&&\nonumber
{\cal
E}={\gamma\over(2\pi)^3}\left(\sum_{B,(B\neq\Sigma)}\int^{k_{F_B}}\d^3k\sqrt{k^2+{M^*_B}^2}+\sum_l\int^{k_{F_l}}\d^3k\sqrt{k^2+{M_l}^2} \right)\\
&&\label{CalEa} +{1\over2}m_\sigma^2\sigma^2 +{1\over2}m_\omega^2\omega^2
+{1\over2}m_{b_3}^2b_3^2,
\end{eqnarray}
and the thermodynamical potential is given by
\begin{eqnarray}
&&\Phi={\gamma\over 2\pi^2}\left(\sum_{B,(B\neq\Sigma)}\int^{k_{F_B}}k^2\d k\left(\sqrt{k^2+{M^*_B}^2}
-(\mu-q_B\lambda)\right)+\int^{k_{F_l}}k^2\d k\left(\sqrt{k^2+{M_l}^2}-\lambda\right)\right)\nonumber\\&&
+{1\over2}m_\sigma^2\sigma^2+{1\over2}m_\omega^2\omega^2+{1\over2}m_{b_3}^2b_3^2, \label{Phi-ea}
\end{eqnarray}
where the Lagrange multiplier $\mu$ controls the baryon density and
$\lambda$ the electrical charge. The $\sigma$ field is determined
from the minimization of $\Phi$ with respect to $\sigma$,
\begin{equation}
  \frac{\partial \Phi}{\partial\sigma}=0.
  \label{sigma}
\end{equation}
The coupling of the scalar meson $\sigma$, which, in the spirit of
SU(3) flavor symmetry, is the same for all quarks, is encapsulated
into the definition of $M^{*}_B$, Eq. (\ref{MB}). The sources of the fields $\omega$ and $b_3$ respectively $\rho_0$, and $\rho_3$
are given by
\begin{eqnarray}
&&\rho_0= {\gamma\over(2\pi)^3}\sum_{B,(B\neq\Sigma)}\zeta_B \int^{k_{FB}}\d^3k,\quad
\rho_3={\gamma\over(2\pi)^3}\sum_{B,(B\neq\Sigma)}
\eta_B\int^{k_{FB}}\d^3k,
\label{rhoa}
\end{eqnarray}
with
\begin{eqnarray}
&&\zeta_P=\zeta_N=1,~\zeta_\Lambda=
1+\delta,~\zeta_{\Xi_0}=\zeta_{\Xi_-}=1+2\delta,
\\\nonumber
&&\eta_P=1,~\eta_N=-1,~\eta_\Lambda=
0,~\eta_{\Xi_0}=1,~\eta_{\Xi_-}=-1.
\end{eqnarray}
The relation between the fields and the respective sources is given by
\begin{equation}\omega={3g^q_\omega\rho_0\over m^2_\omega},\quad b_3={g^q_{b_{3}}\rho_3\over m^2_{b_{3}}}.
\label{omegaa}
\end{equation}


We start by fixing the free parameter $\kappa$ of the Bogoliubov model. This is obtained by fitting the nucleon mass M = 939 MeV.
Next, the desired values of the neutron effective mass
$M^{*}/M=0.773$, 
nuclear matter binding energy $E_{B}=\epsilon/\rho_{B}-M_{N}=-15.7$MeV,
the incompressibility $K =315.0$ MeV, in agreement with the range  of
values proposed in \cite{stone2014},  and the nucleon radius
$R_B=0.1163$ fm at saturation density, $\rho_{B}= 0.145$fm$^{-3}$, are obtained by setting $g^q_{\sigma}= 4.0539996$ and $3g^q_{\omega}= g^q_{\omega N} = 9.2474196$. The coupling constant $g^q_{b_{3}}=3.9532889$
is fixed in order to have the symmetry energy coefficient $a_4=29$MeV
and the symmetry energy slope $L=79.45$ MeV, at saturation
density. The value we consider for $L$ is well inside the 
range  of values
obtained in \cite{oertel2018} from a huge number  of experimental
data and astrophysical observations, $L= 58.7 \pm28.1 $ MeV.
We have chosen a slightly low saturation density in order that the
model produces reasonable values of incompressibility $K$.


There is an appreciable mass difference between the hyperons $\Lambda$
and $\Sigma$, which, according to~\cite{isgur,Le Yaouanc,capstickroberts}
is due to an hyperfine splitting. Moreover, it should be kept in mind that the $SU(2)$ symmetry is a very important one.
The hyperon $\Lambda$ is an isosinglet; the nucleon and the $\Xi$ are isodoublets; the $\Sigma$ is an isotriplet.
Besides, it is known that the $\Sigma$-nucleus potential  in symmetric
nuclear matter seems to be repulsive \cite{meissner,gal}.
Our nucleon model does not take into account the mechanism responsible for
the above mentioned hyperfine splitting,  the $\Sigma$ and
$\Lambda$ hyperons  are degenerate,  and
besides also leads to an attractive optical potential for the $\Sigma$.
We overcome this problem by omitting the $\Sigma$ in sums over $B$, as explicitly indicated in
(\ref{rhoa}), and in analogous sums in the sequel.  We are only performing sums over baryons which are
either isosinglets or isodoublets.
The omission of the   $\Sigma$-hyperon is in
accordance with  the general result obtained  when a  repulsive
$\Sigma$-potential in symmetric matter at saturation density of the order of
30 MeV is considered \cite{chatterjee1, chatterjee2, fortin2016}:
$\Sigma$-hyperons are not present inside neutron stars.

\begin{figure}[th]

\includegraphics[width=0.85\linewidth,angle=0]
{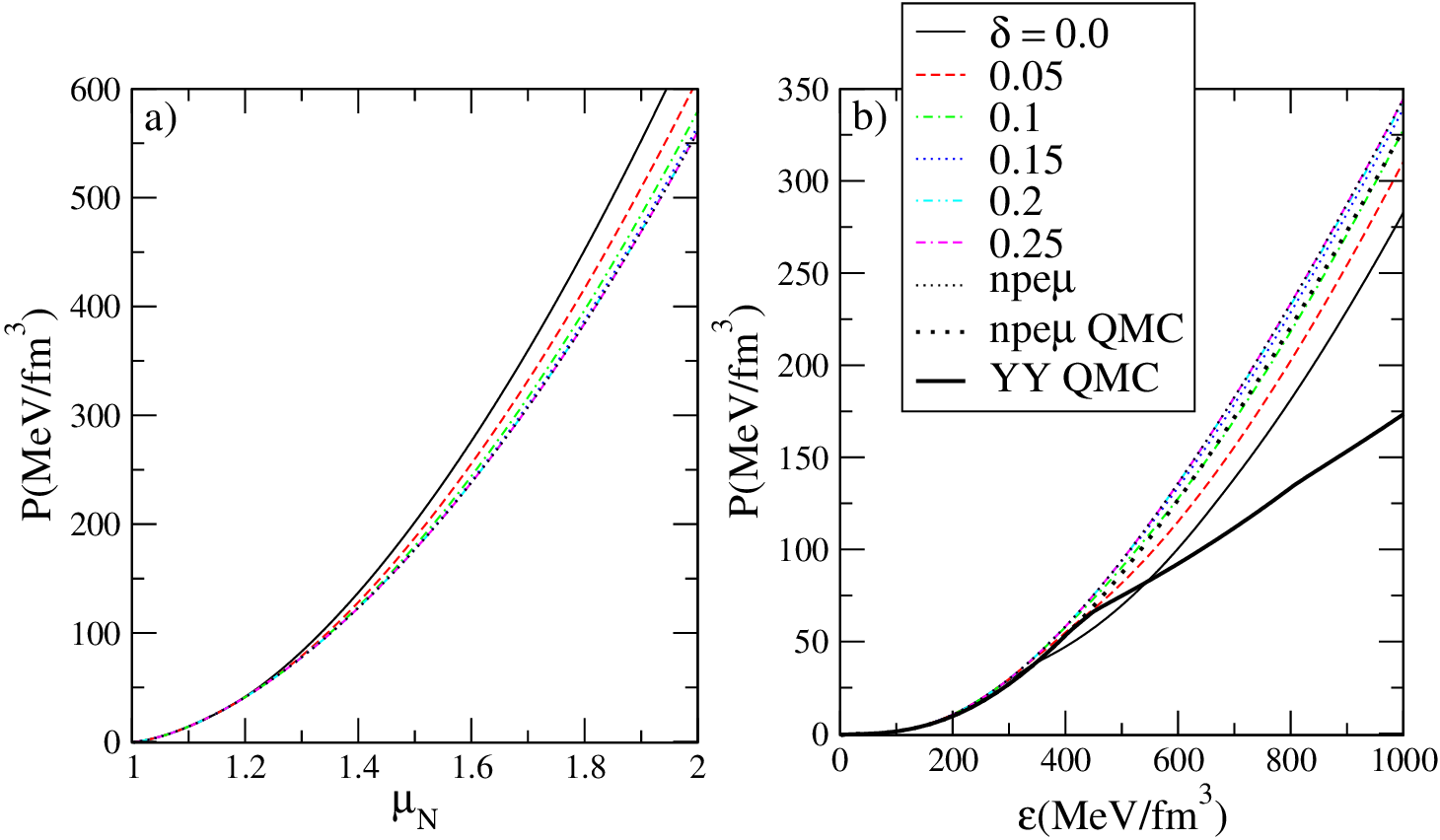}            
\caption{Pressure versus neutron chemical potential comparing neutron, proton, leptonic
  matter with hyperonic matter for $\delta=0.0,\,
  0.05,\,0.1,\,0.15,\,0.2$ and 0.25 (left panel), and the pressure versus energy
  density   for  $\beta$-equilibrium
  nucleonic and  hyperonic matter  compared with the corresponding QMC EoS obtained in
  \cite{panda12}.  The thick lines have been obtained with QMC models
  \cite{panda12} and all the thin lines represent EoS obtained for the
  Bogoliubov-QMC model with and without hyperons.}
\label{eos}
\end{figure}

Minimization of $\Phi$ with respect to $k_{F_B}$ leads to
\begin{equation}\label{Fenergya}
\sqrt{k_{FB}^2+M^{*2}_B}
+3g_\omega^q\omega\zeta_B+g^q_{b_3} b_3\eta_B 
=\mu-q_B\lambda.
\end{equation}

The quantity $\mu-q_B\lambda$ is usually referred to as the chemical potential of baryon $B$.
Minimization of $\Phi$ with respect to $k_{F_e}$ leads to
\begin{equation}\label{eFEa}
\sqrt{k_{Fe}^2+M_e^2}=\lambda,\end{equation}
so the Lagrange multiplier $\lambda$ is usually called the electron Fermi energy.

\begin{figure}[th]
\centering
\includegraphics[width=0.7\linewidth,angle=0]{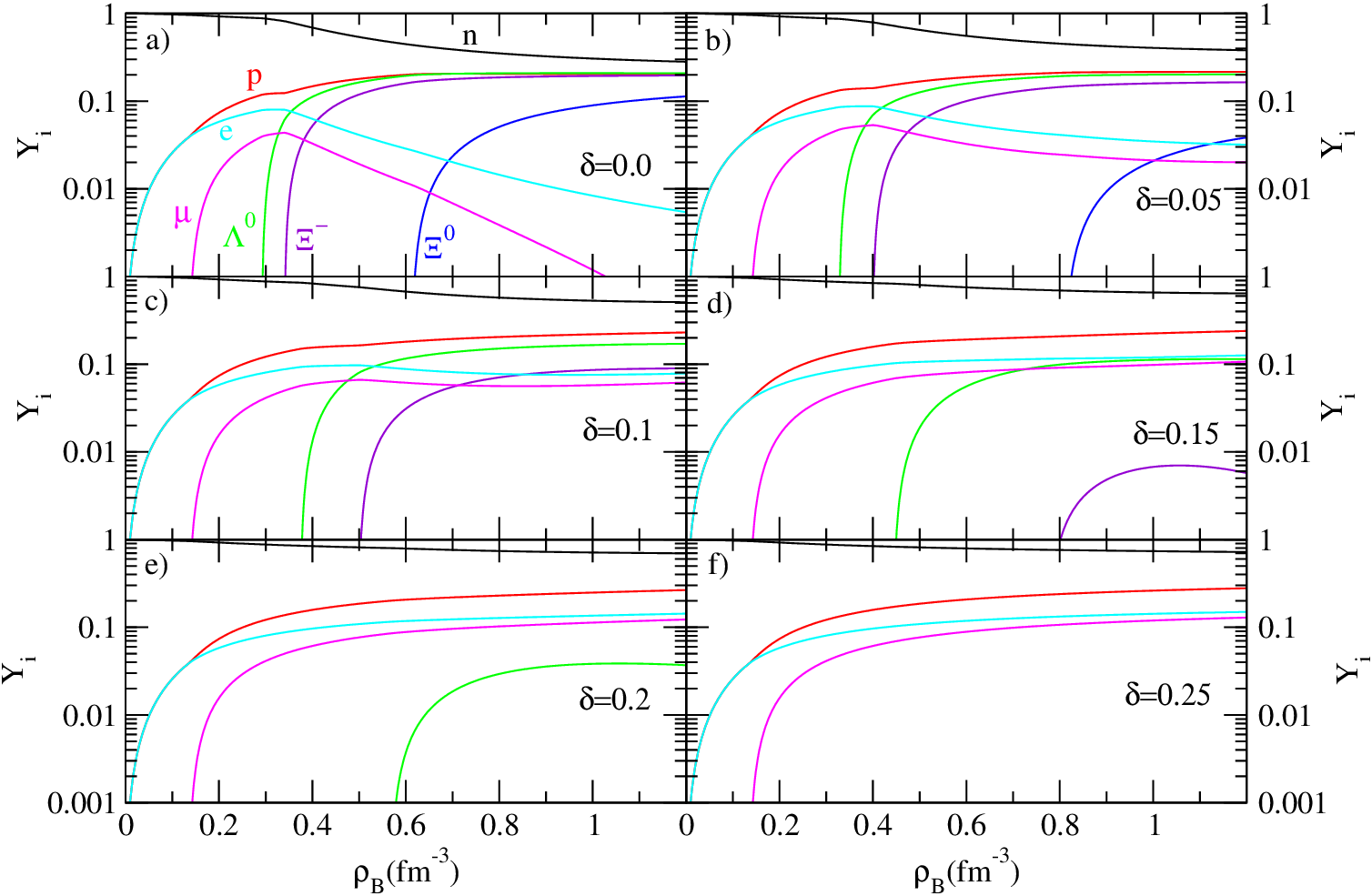} 
\caption{
Baryonic and leptonic particle fractions as a function of the baryonic
density, for several values of the parameter $\delta$. For
$\delta=0.25$ the onset of hyperons is shifted to densities above 1.2
fm$^{-3}$. The central baryonic density lies between 0.9 and 1.1
fm$^{-3}$ depending on the hyperonic content.
}
\label{fraction}
\end{figure}

 Explicitly, for $N,~\Lambda,~\Xi,$ 
 (\ref{Fenergya}) reduces to
\begin{eqnarray*}&&\nonumber
\sqrt{k_{F_N}^2+M^{*2}_N}
+3 g_\omega^q\omega+g^q_{b_3} b_3\eta_N
=\mu-q_N\lambda
,\\
&&\nonumber
\sqrt{k_{F_\Lambda}^2+M^{*2}_\Lambda}
+3 g_\omega^q(1+\delta)\omega
=\mu
,\\
&&
\label{FenergLX}
\sqrt{k_{F_\Xi}^2+M^{*2}_\Xi}
+3 g^q_\omega(1+2\delta)\omega+g^q_{b_3} b_3\eta_\Xi 
=\mu-q_\Xi\lambda.
\end{eqnarray*}

Then, according to the prescription of \cite{glendenning}, we have
\begin{eqnarray*}
&&U_\Lambda:=M^{*}_\Lambda-M_\Lambda+3g^q_\omega(1+\delta)\omega
,\\
&&U_\Xi:=M^{*}_\Xi-M_\Xi+3g_\omega^q(1+2\delta)\omega,
\end{eqnarray*}
and it is possible to fix the coupling to the quark $s$ in such a way that a reasonable $U_\Lambda$,
is obtained. 
We find that a small change of $\delta$ leads to big changes of $U_\Lambda$ and $U_\Xi$. However, the
EoS is almost insensitive to the value of $\delta$ for a wide range of values of $U_\Lambda$ around the
proper one.
This model predicts a competition
between negatively charged hyperons and leptons. This is
natural in view of Bodmer-Witten's Conjecture \cite{bodmer,witten}, according to which the groundstate of baryonic matter at high densities
should involve only quarks $u,\, d,\, s,$ without leptons.

The pressure $P$ obtained for different values of $\delta$  is
  shown in Fig.~\ref{eos} as a function of the baryonic chemical
  potential $\mu_N$ (left panel) and the energy density (right
  panel). In the right panel we also include the EoS obtained with the
  QMC model \cite{panda12} for comparison. The inclusion of hyperons
  softens the EoS as expected, corresponding to a larger pressure for a
  given chemical potential. When the $\delta$ parameter is turned on
  the EoS gets stiffer and, for the range of the chemical potential
  shown, the EoS with $\delta\ge 0.2$ almost coincides with the
  nucleonic EoS. The QMC EoS softens strongly above 3$\rho_0$ and becomes
  much softer than the Bogoliubov-QMC EoS, even with $\delta=0$.

In Fig. \ref{fraction}, the fractions of baryons and leptons for
  $\beta$-equilibrium matter are represented as a function of density
  for different values of the parameter $\delta$. For $\delta=0$ the
  onset of hyperons occurs for $\rho=2\rho_0$, and the first hyperon
  to set in is the $\Lambda$ and at a slightly larger density the $\Xi^-$. A
  finite $\delta$ pushes the onset of these particles to larger
  densities and for $\delta=0.2$ the onset of $\Lambda$ occurs at
  $\sim 4
  \rho_0$ and the $\Xi^-$ above 8$\rho_0$. Besides the fraction of
  $\Lambda$s in this last scenario never goes above 4\% while for
  $\delta=0$ it reachs 20\%. 

\begin{figure}[th]
\begin{tabular}{cc}
\includegraphics[width=0.46\linewidth,angle=0]{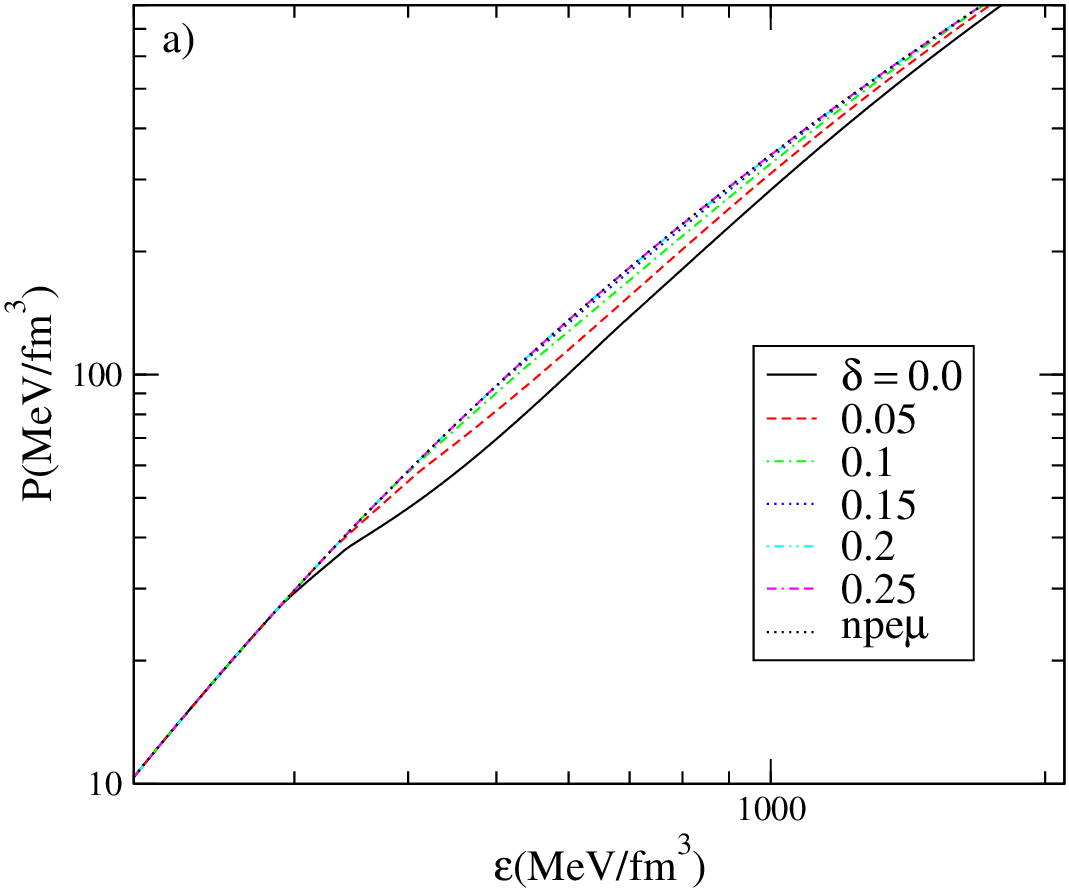} & \includegraphics[width=0.54\linewidth,angle=0]{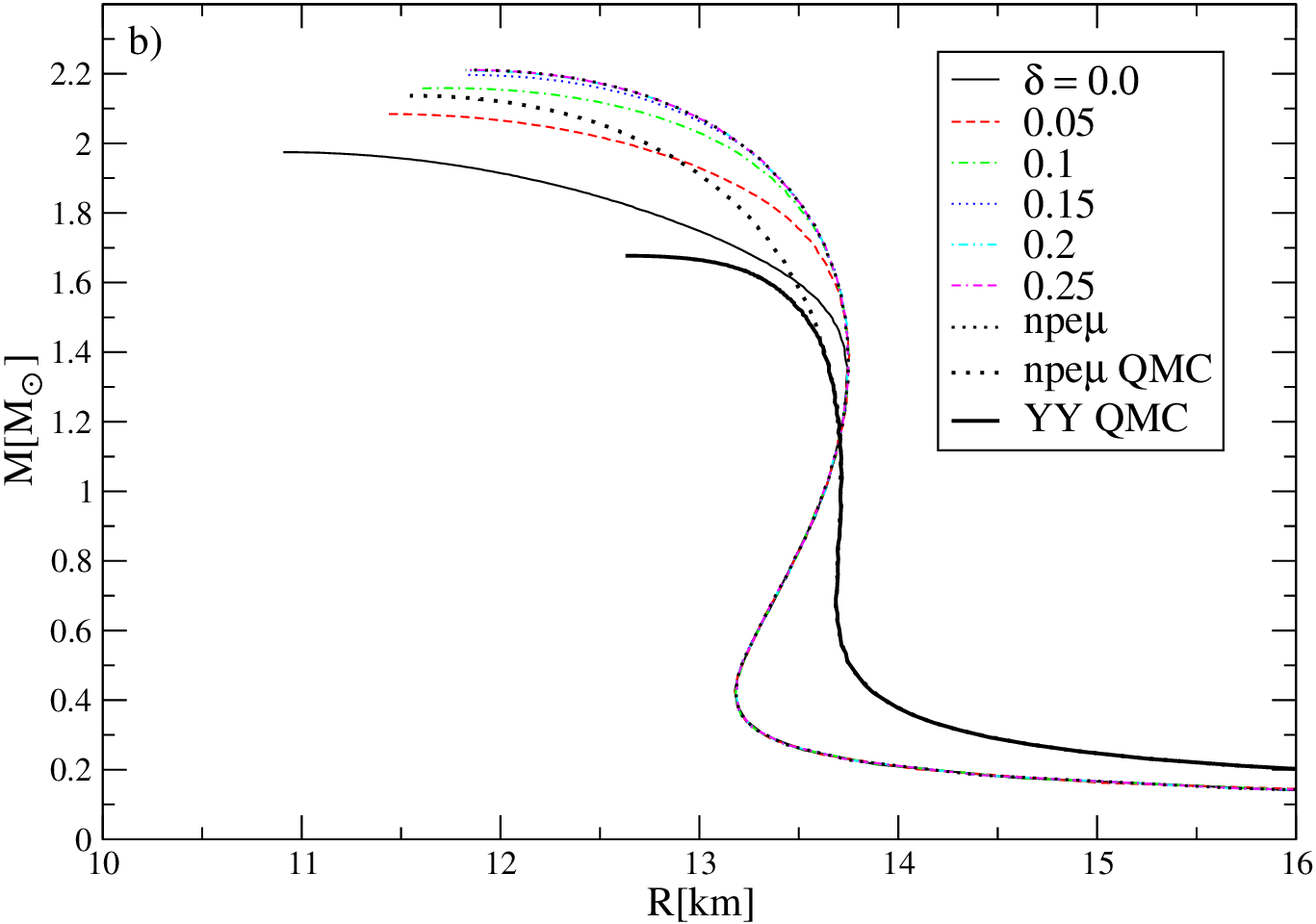}
\end{tabular}
\caption{ EoS (left) and mass-radius curves  obtained from the integration of the TOV
  equations (right), for different values of the $\delta$ parameter. The
  curves stop at the maximum mass configuration. The family of stars
  for nucleonic stars constituted by $npe\mu$ matter is also
  represented. Also shown are the mass-radius curves obtained with the
  QMC model in \cite{panda12} for comparison (thick full and dotted lines).}
\label{mr}
\end{figure}

\begin{table*}[!htb]
\caption{Properties of the stable neutron star with maximum mass, for
  several values of $\delta$, $M_{max}$, $M^b_{max}$, $R$, $E_{0}$,
  $\rho^c$, $R_{1.4}$, $R_{1.6}$,  $U_\Lambda(\rho_0)$ and
  $U_\Xi(\rho_0)$ are respectively, the gravitational and baryonic
  masses, the star radius, the central energy density, the central
  baryonic density, the radius of neutrons star calculated for ${1.4
    M_{\odot}}$ and ${1.6 M_{\odot}}$, and the optical potentials for
  a $\Lambda$ and $\Xi$-hyperon in symmetric nuclear matter at saturation.}
\label{table}
\begin{ruledtabular}
\begin{tabular}{ccccccc c cc c}
$\delta$ & $M_{max} $ & $M^b_{max}$& R &
                                                                  $E_{0}$
  & $u^ c=\rho^c/\rho_0$ & $R_{1.4}$ & $R_{1.6}$&
                                                          $U_\Lambda(\rho_0)$&$U_\Sigma(\rho_0)$&$U_\Xi(\rho_0)$
                                                        \\
& $[M_\odot]$ &$[M_\odot]$& [km]&  [{fm}$^{-4}$] & & [km]  &[km]&  [MeV] & [MeV] & [MeV]\\
\hline
$0.0$    & 1.97 & 2.28 & 10.91 & 7.25 & 7.674 &  13.731  & 13.492 & -75.34&& -93.99 \\
$0.02$   & 2.02 & 2.34 & 11.16 & 6.85 & 7.293 &  13.740  & 13.624
                                                &-72.23 &&-87.77 \\
$0.05$   & 2.08 & 2.43 & 11.42 & 6.43 & 6.882 &  13.752  & 13.680   &-67.57&& -78.45\\
$0.1$    & 2.16 & 2.53 & 11.73 & 5.97 & 6.429 &  13.750  & 13.693
                                                &-59.80 && -62.91\\
$0.15$   & 2.20 & 2.58 & 11.83 & 5.84 & 6.285 &  13.746  & 13.698 &-52.03&&  -47.37 \\
$0.2$    & 2.21 & 2.60 & 11.84 & 5.81 & 6.256 &  13.748  & 13.697
                                                &-44.26&&   -31.83\\
$0.25$   & 2.21 & 2.60 & 11.84 & 5.82 & 6.255 &  13.746  & 13.696 &-36.49&&-16.29\\
npe$\mu$ & 2.21 & 2.60 & 11.84 & 5.84 & 6.272 &  13.746  & 13.696 & &&\\
\hline
npe$\mu$ QMC &  2.14 & 2.50  & 11.54 & 6.21 & 6.434 & 13.628  & 13.485 & &&\\
YY QMC & 1.68 & 1.88 & 12.63 & 4.57 & 5.345 & 13.624  & 13.344 &-28.0 &+30.0 & -18.0\\
\end{tabular}
\end{ruledtabular}
\end{table*}


In order to study the structure of neutron stars described by the
present model we have integrated  the
Tolman-Oppenheimer-Volkov equations  for spherical  stars in
equilibrium \cite{tolman,tov}. The complete  EoS  was obtained
matching the Baym-Pethcik-Sutherland EoS for the outer crust \cite{bps}, and the inner crust obtained within a Thomas Fermi description of the non-homogeneous
matter for the NL3$\omega\rho$ model with the symmetry energy slope at
saturation equal to 77 MeV \cite{pais2016}, to the core  EoS.
It has been discussed in \cite{fortin2016} that a non-unified EoS,
i.e. a neutron star EoS constituted by a crust and core EoS obtained
from different models gives rise to an uncertainty on the radius of
low mass stars. For the crust, two distinct contributions are
included: the outer crust and the inner crust. While the first one is
relatively well constrained,  the second one is clearly dependent on the
underlying model used to describe the star.
In the present work we consider the conclusion drawn in
\cite{pais2016} where it was shown that taking an inner crust
EoS from a model with a similar 
dependence of the symmetry energy on the density would predict realistic
radii. The complete EoS,  that has been used to integrate the
TOV equations, is plotted in  the left panel of Fig. \ref{mr} in
a log-log scale so that it is clearly seen that the  crust-core transition is
smooth. In Table \ref{table} several star properties are given, including
the maximum gravitational  mass $M_{max}$, and corresponding  baryonic
mass  $M^b_{max}$, radius $R$, energy density  and baryonic density at the centre $E_{0}$,
 and  $\rho^c$,  the radius of 1.4$M_\odot$ and 1.6$M_\odot$ stars,
 $R_{1.4}$ and $R_{1.6}$,  and the hyperon potentials in symmetric
 nuclear matter at saturation $U_\Lambda(\rho_0)$ and
  $U_\Xi(\rho_0)$.

For $\delta=0$, we find that the EoS does not describe stars with
masses above 1.92 solar masses.
For $\delta\geq0.2$, the EoS and the curve mass
vs. radius are almost insensitive to the value of $\delta$.  The onset
of hyperons  occurs at a density above $\sim 0.6$ fm$^{-3}$ and the hyperon
fraction is too small. Let us point out that we obtain reasonable
values for the hyperon-potentials in symmetric nuclear matter for
$\delta\sim 0.25$, see Table \ref{table}.  With this value of $\delta$ no hyperons will
appear inside neutron stars. A similar conclusion was obtained by
\cite{pederiva2015} within a microscopic approach that includes three
body contributions of the form $NNY$. Under   these results,  two
solar  mass  stars will not contain hyperons
because they will set in at densities of the order of the neturon star
central density or above. At this point let us add a comment
 on the hyperon content within other QMC models. In
 refs. \cite{qmcPanda1,panda12} hyperons have been included in the QMC
 model \cite{guichon1}. Taking different choices for
the meson-hyperon couplings, the onset of hyperons occurs below 3
$\rho_0$. Due to the softening the EoS with the hyperon onset the
maximum star mass decreases from $\approx 2.2 M_\odot$ to $\approx 1.8
M_\odot$, see Fig. \ref{mr}b) where the nucleonic M/R curve is plotted for the
nucleonic EoS (thick dotted line)  and the hyperonic EoS (thick full line). Including a non-linear $\omega\rho$ term that allows to
soften the density dependence of the symmetry energy pushes the
hyperon onset to densities above 3
$\rho_0$ and increases the maximum star mass to $\approx 1.9
M_\odot$.  Considering the improved QMC model \cite{guichon2},    
Stone {\it et al.} have also included hyperons \cite{stone07,stone18}. This version of the model, which
takes into account the color hyperfine interaction and the scalar
polarisability of the baryons and, besides considers  the Hartree–Fock
approximation,  predicts that hyperons set in close to 4
$\rho_0$ and maximum star masses lie in the interval $\approx 1.9-2 M_\odot$. Even
though in smaller amount hyperons do set in inside neutron stars.
Within the  MQMC model
\cite{barik3} the EoS is quite sensitive to the quark mass and only taking a quark
  mass above 200 MeV, is it possible to describe neutron
  stars with masses above $1.9$
  $M_\odot$ containing small amounts of hyperons in their core.

The canonical star with a mass 1.4 $M_\odot$ has a radius  of the order
of 13.7 km, well within the  values obtained by NICER
\cite{nicer1,nicer2}  and other observations \cite{haensel}, and within or 
just slightly above the predition obtained from terrestrial data
\cite{li2006}, or the gravitational wave GW170817
\cite{abbott17,abbott18}  detected by LIGO/Virgo from a neutron neutron star merger \cite{de2018,fattoyev18,tuhin}.  We have calculated the  tidal
deformability  of a canonical star  with a mass
1.4$M_\odot$ according to \cite{hinderer}.
The result obtained was $\Lambda_{1.4}=936 - 954$ depending on the
hyperon content, well  above  the
prediction of \cite{abbott18} $70<\Lambda_{1.4}<580$, which however
was determined from  a set of models that do not necessarily describe two
solar-mass stars. These high values of  $\Lambda_{1.4}$ may indicate
that  the symmetry energy is too stiff as discussed in
\cite{fattoyev18,veronica}, and the  inclusion of a non-linear $\omega-\rho$
term will soften the symmetry energy at high densities, and decrease
the value of $\Lambda_{1.4}$.

Within the present model we are able to describe  NS as massive as
the pulsar MSP J0740+6620 \cite{cromartie2020}, in particular, if
we constrain the  optical potential   of the $\Lambda$-hyperon in
symmetric matter to experimental values.
Only the hyperon fraction of baryonic matter and the value of the optical potential are sensitive to the precise value of $\delta$, for $\delta\geq0.2$.

\section{Conclusions}

In the present study we have developed a QMC model based in the
Boguliubov quark model. The nucleons interact via
the exchange of a scalar-isoscalar meson, a vector-isoscalar meson and a
vector-isovector meson. The  baryon mass is derived from the baryon energy
 which  includes $u$, $d$ and $s$ quarks. The parameters
introduced at this level are chosen so that the vacuum constituent
quarks masses are reproduced. Hadronic matter is described by
introducing a   vector-isoscalar $\omega$-meson, which also includes
a $\bar s s$ content, and a  vector-isovector $b_3$-meson.
In order to satisfy constraints imposed by neutron stars and
hypernuclei it is shown that the coupling of the $\omega$-meson
to the $s$-quark must  be more repulsive than its coupling to the $u$, and
$d$-quarks, and a parameter that  takes this aspect into account has
to be introduced, so that SU(3) symmetry is broken. 

The couplings of the mesons to the nucleons were fixed so that nuclear
matter properties, binding energy, saturation density,
incompressibility,  symmetry energy and its slope at saturation, are
adequately described. Once these parameters are fixed, only the 
parameter that defines how repulsive is the coupling of the
$\omega$-meson to hyperons, remains to be fixed. Taking the optical potential of the
$\Lambda$-hyperon of the order of $-30$ MeV as discussed in
\cite{gal,shen06,vidana15}, no hyperons will be present inside a
two-solar mass. A similar conclusion has been drawn in
\cite{pederiva2015} where, within  an auxiliary field diffusion Monte Carlo
algorithm, it was shown that  the three-body hyperon-nucleon
interaction has an important role in softening the EoS at large
densities. Using experimental separation energies of medium-light
hypernuclei to constrain the $\Lambda NN$ force, they have shown that
the onset of hyperons will occur above 0.56 fm$^{-3}$, and concluded
that with the presently available experimental energies of
$\Lambda$-hypernuclei it is not  possible to draw a conclusive
statement concerning the presence of hyperons inside neutron stars.


The present model predicts for the canonical neutron star a radius
that is compatible with observations and predictionns from the
analysis of the GW170817 detection. The tidal deformability, is
however, too large, and this may indicate that the symmetry energy is
too stiff. A softer symmetry energy may be generated with the inclusion of a non-linear
$\omega-\rho$ term  in the model \cite{horowitz01}.

\begin{acknowledgments}
This work was partially supported by national funds from  FCT  (Fundação  para  a  Ciência e a  Tecnologia, I.P, Portugal) under the Projects No. UID/FIS/04564/2019, No. UID/04564/2020, and POCI-01-0145-FEDER-029912 with financial support from  POCI, in its FEDER  component, and  by the FCT/MCTES budget through national funds (OE). 
\end{acknowledgments}

\end{document}